\documentclass[pdflatex,sn-mathphys-num]{sn-jnl}

\usepackage{graphicx}%
\usepackage{multirow}%
\usepackage{amsmath,amssymb,amsfonts}%
\usepackage{amsthm}%
\usepackage{mathrsfs}%
\usepackage[title]{appendix}%
\usepackage{xcolor}%
\usepackage{textcomp}%
\usepackage{hyperref}
\usepackage{manyfoot}%
\usepackage{bm}
\usepackage{booktabs}%
\usepackage{algorithmicx}%
\usepackage{listings}%
\usepackage{braket}
\usepackage[ruled,vlined]{algorithm2e}

\theoremstyle{thmstyleone}%
\theoremstyle{thmstyletwo}%

\theoremstyle{thmstylethree}%

\begin{document}

\title[]{Solving Differential Equation with Quantum-Circuit Enhanced Physics-Informed Neural Networks }


\author[]{\fnm{Rachana} \sur{Soni}}\email{rachanasoni007@gmail.com}

\affil{\orgdiv{School of Computer Science Engineering and Technology}, \orgname{Bennett University}, \orgaddress{\street{TechZone 2}, \city{Greater Noida}, \postcode{201310}, \state{U.P.}, \country{India}}}



\abstract{

I present a simple hybrid framework that combines physics-informed neural networks
(PINNs) with features generated from small quantum circuits. As a proof of concept,
the first-order equation $\tfrac{dy}{dx}+2y=0$ with $y(0)=1$ is solved by feeding
quantum measurement probabilities into the neural model. The architecture enforces
the initial condition exactly, and training is guided by the ODE residual loss.
Numerical results show that the hybrid model reproduces the analytical solution
$e^{-2x}$, illustrating the potential of quantum-enhanced PINNs for differential
equation solving.}




\maketitle

\section{Introduction}

Differential equations play a central role in modeling physical, biological, and
engineering systems. Traditional numerical solvers, while well established, can become
computationally demanding in high dimensions or when equations are only partially known.
Physics-informed neural networks (PINNs) have emerged as a data-driven alternative,
where the governing equation is embedded into the loss function of a neural network.
This framework allows approximate solutions to be obtained without explicitly constructing
finite-difference or finite-element discretizations.

At the same time, quantum computing offers new ways of encoding and processing information.
In particular, the statistics generated by shallow quantum circuits can be viewed as
nonlinear feature maps that are difficult to reproduce classically. These quantum features
may enrich the representation power of machine-learning models even on near-term devices.
Recent studies have explored their use in supervised and generative tasks, but their
application to scientific computing remains largely unexplored.

In this work, I investigate the use of quantum-circuit features within a PINN framework
for solving differential equations. As a proof of concept, we focus on the linear
ordinary differential equation $\tfrac{dy}{dx}+2y=0$ with initial condition $y(0)=1$.
A simple two-qubit circuit is employed to generate measurement probabilities, which are
concatenated with the collocation points and used as inputs to the neural network.
The architecture is constructed to satisfy the initial condition exactly, while the
training process minimizes the ODE residual loss. The numerical results indicate that
the hybrid model can approximate the analytical solution $e^{-2x}$, suggesting that
quantum-enhanced features may serve as a useful ingredient in neural solvers for
differential equations.

Recent years have witnessed an increasing interest in leveraging quantum
mechanics as a computational resource.. Foundational theory of quantum
 mechanics \cite{griffiths2018introduction,teschl2014quantum,nielsen2010quantum}
 establish the mathematical models.

Parallel to these advances, physics-informed and operator-based neural
architectures have emerged as powerful tools for solving differential equations
in scientific computing. Physics-informed neural networks (PINNs) incorporate
governing equations directly into the loss function
\cite{raissi2019physics,karniadakis2021physics}, while neural operators learn
maps between function spaces and generalize across problem instances
\cite{kovachki2023neural,li2020fourier}. Applications to high-dimensional PDEs
\cite{han2018solving} and quantum systems
\cite{mills2017deep,carleo2017solving,carrasquilla2017machine} further
demonstrate their versatility. By integrating quantum circuit generated feature with PINNs, one can bridge these two lines of research,
opening a pathway for hybrid quantum--classical solvers that address
differential equations from a new perspective.

\subsection*{Significance and Motivation}
The significance of this study lies in bridging physics-informed neural networks (PINNs)
with quantum feature maps. While PINNs have proven effective for solving ordinary and
partial differential equations, their accuracy can be limited by the quality of the input
features. On the other hand, parameterized quantum circuits naturally generate structured,
nonlinear features that are difficult to reproduce classically. Integrating such features
into the PINN framework represents a novel direction for scientific machine learning.

The need for this work arises from a gap in the literature. Although quantum circuits have
been employed in classification and generative modeling tasks, their role in solving
differential equations has remained largely unexplored. Existing PINN approaches rely
entirely on classical features, and little attention has been paid to hybrid models that
leverage quantum statistics. By demonstrating the feasibility of solving the simple ODE
$\tfrac{dy}{dx}+2y=0$ with quantum-circuit features, this work provides an initial step
towards quantum-enhanced solvers for more complex systems of scientific interest.

\section{Mathematical Framework}

We consider the first-order linear ODE
\begin{equation}
\frac{dy}{dx} + 2y = 0, \qquad y(0) = 1,
\label{eq:ode}
\end{equation}
whose analytical solution is
\begin{equation}
y(x) = e^{-2x}.
\end{equation}

\subsection{Quantum-Circuit Feature Encoding}
For each grid point $x \in [0,1]$, we define a 2-qubit parameterized circuit
\begin{equation}
U(t) = \Big( Rx(-2t) \otimes Rx(-2t) \Big) \, \big(H \otimes I\big),
\end{equation}
with evolution parameter $t = x$.  
The quantum state is
\begin{equation}
|\psi(x)\rangle = U(t) |00\rangle.
\end{equation}

Measuring $|\psi(x)\rangle$ in the computational basis yields a probability vector
\begin{equation}
p_Z(x) = \big[ P_Z(00), \, P_Z(01), \, P_Z(10), \, P_Z(11) \big].
\end{equation}

The quantum feature vector is then
\begin{equation}
q(x) =
\begin{cases}
p_Z(x), & \text{(4-dimensional)} .
\end{cases}
\end{equation}
\begin{figure}[h!]
    \centering
    \includegraphics[width=1\linewidth]{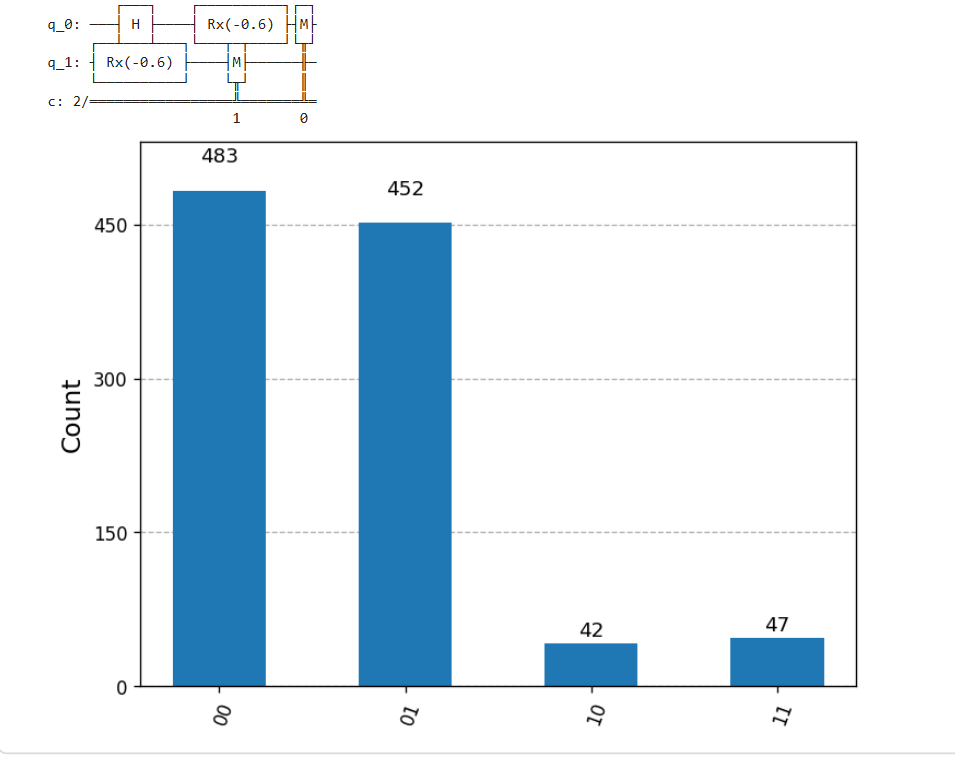}
   \caption{Histogram of measurement outcomes from the two-qubit quantum circuit 
    used for feature generation. The probabilities of the computational basis states 
    \{00, 01, 10, 11\} are estimated from repeated circuit executions (shots). 
    These distributions provide the quantum features that are integrated into the 
    neural ODE solver.}
    \label{fig:hist_qc}
\end{figure}
\subsection{Neural Network Ansatz with Hard Initial Condition}
We define the neural ansatz as
\begin{equation}
\hat y(x) = 1 + x \, g_\theta\!\big( [x,\, q(x)] \big),
\end{equation}
where $g_\theta$ is a feed-forward neural network with trainable parameters $\theta$.
This construction enforces the initial condition $\hat y(0) = 1$ exactly.

\subsection{Physics-Informed Loss}
The residual of \eqref{eq:ode} at each collocation point $x_i$ is
\begin{equation}
r(x_i; \theta) = \frac{d \hat y}{dx}(x_i) + 2 \hat y(x_i).
\end{equation}
The physics-informed loss is
\begin{equation}
\mathcal{L}_{\text{ODE}}(\theta) = \frac{1}{N} \sum_{i=1}^N r(x_i;\theta)^2.
\end{equation}

\subsection{Training Objective}
The parameters $\theta$ are optimized by minimizing
\begin{equation}
\mathcal{L}(\theta) = \mathcal{L}_{\text{ODE}}(\theta),
\end{equation}
using stochastic gradient descent (Adam optimizer).  
The trained model $\hat y(x)$ then approximates the analytical solution $e^{-2x}$.

\begin{algorithm}[H]
\caption{Quantum Feature Generation }
\KwIn{Collocation points $x_1,\dots,x_N \in [0,1]$, shots $S$}
\KwOut{Feature matrix $Q \in \mathbb{R}^{N\times 4}$ with rows $q(x_i)$}

\For{$i \gets 1$ \KwTo $N$}{
  Set $t \leftarrow x_i$ \tcp*{parameterize by input}
  Prepare 2-qubit circuit $U(t)$: apply $H$ on $q_0$; apply $Rx(-2t)$ on $q_0,q_1$\;
  Measure both qubits in the Z basis (shots $S$) to get counts over $\{00,01,10,11\}$\;
  Convert counts to probabilities $p_Z(x_i) = [P(00),P(01),P(10),P(11)]$\;
  Set $q(x_i) \leftarrow p_Z(x_i)$ \tcp*{Z-only, 4-dim}
}
Stack $q(x_i)$ as rows to form $Q$; standardize columns $(Q - \mu)/\sigma$\;
\end{algorithm}
\begin{algorithm}[H]
\caption{PINN for $y'+2y=0$ with Quantum Features (Hard IC)}
\KwIn{Points $x_1,\dots,x_N$, feature matrix $Q \in \mathbb{R}^{N\times 4}$}
\KwOut{Trained network $\hat y(x)$}

Define MLP $g_\theta:\mathbb{R}^{1+4}\!\to\!\mathbb{R}$; input $z(x)=[x,\,q(x)]$\;
Enforce hard initial condition via $\hat y(x)=1 + x\, g_\theta(z(x))$ so $\hat y(0)=1$\;

\BlankLine
\textbf{Physics loss:}
For each $x_i$, compute $\hat y'(x_i)$ by automatic differentiation w.r.t.\ $x$ and form
$r(x_i)=\hat y'(x_i)+2\,\hat y(x_i)$; set
$\displaystyle \mathcal{L}_{\text{ODE}}(\theta)=\frac{1}{N}\sum_{i=1}^N r(x_i)^2$.

\BlankLine
\textbf{Training:}
Initialize $\theta$; \For{$\text{epoch}=1$ \KwTo $T$}{
  Evaluate $\mathcal{L}_{\text{ODE}}(\theta)$ on $\{x_i,Q_i\}$\;
  Update $\theta \leftarrow \theta - \eta\,\nabla_\theta \mathcal{L}_{\text{ODE}}(\theta)$ (Adam)\;
}
Return $\hat y(x)=1+x\,g_\theta([x,q(x)])$.
\end{algorithm}

\section{Results}
\begin{figure}[h!]
    \centering
    \includegraphics[width=1\linewidth]{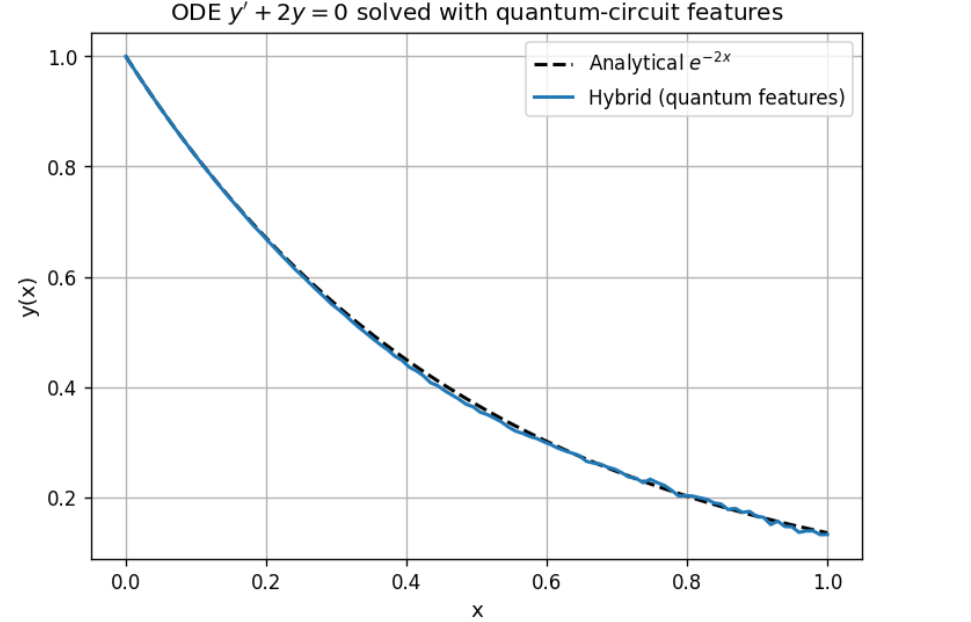}
    \caption{Comparison of the analytical solution $y(x) = e^{-2x}$ (dashed line) 
    with the hybrid quantum--classical prediction (solid line) obtained using 
    quantum-circuit features. The close agreement in the region $x \in [0,0.5]$ 
    indicates that the solver accurately captures the exponential decay. 
    Small deviations at larger $x$ values are due to finite sampling noise and 
    limited training, yet the overall trend demonstrates the effectiveness of 
    quantum-generated features in physics-informed ODE solvers.}
    \label{fig:ode_qnn}
\end{figure}

The hybrid solver successfully reproduced the analytical solution
$y(x)=e^{-2x}$ using  quantum features within the PINN framework. As
shown in Fig.~\ref{fig:ode_qnn}, the predicted curve overlaps the ground truth
with only minor deviations at larger $x$ values, confirming the effectiveness of
the approach.
\section{Conclusion}
This study presented a hybrid framework where quantum-circuit statistics were
embedded as features into a physics-informed neural network for solving an
ordinary differential equation. The results demonstrate that quantum-generated
features can effectively enrich the solution space, allowing the hybrid solver to
closely approximate the analytical solution. While the present work is a proof of
concept on a simple ODE, it highlights the potential of integrating  circuit-based features into scientific machine learning. Extending this approach
to more complex equations and higher-dimensional systems offers a promising
direction for future research.

\nocite{*} 
\bibliography{reference}

\end{document}